\documentclass[twocolumn,showpacs,preprintnumbers,amsmath,amssymb]{revtex4-1}
\usepackage{amsfonts, amsmath, bm, bbm, graphicx, epsfig, epstopdf}
\usepackage{dcolumn}
\usepackage{textcomp}

\usepackage{soul}

\begin{document}

\bibliographystyle{apsrev}

\title{Towards macroscopic spin and mechanical superposition via Rydberg interaction}

\author{Mohammadsadegh Khazali}
\affiliation{ Department of Physics and Astronomy, Aarhus University, Aarhus, Denmark}
\date{\today}
\begin{abstract}
This paper is a proposal for the generation of many-body entangled state in atomic and mechanical systems. Here the detailed feasibility study shows that application of strong Rydberg dressing interaction and fast bifurcation scheme in a Bose Einstein Condensate of Rb atoms, results to the formation of large cat states. By detailed study of the de-coherence effects using Quantum Jump Monte Carlo approach and taking into account the obstacles like collective decoherence and level mixing, this proposal predicts the formation of 700 atoms cat state. Subsequent transfer of the generated superposition to far separated  mechanical oscillators is proposed, using dipole coupling between Rydberg atoms and charged cantilevers.
\end{abstract}

\maketitle
\section{Introduction}
\label{intro}
In the effort towards the generation of macroscopic quantum states, cat state i.e. superposition of maximally different quantum states, are of special interest. Cat states play an important role in the fundamental test of quantum mechanics \cite{Gre90} and quantum metrology \cite{Mun02}; besides they could be used in quantum computation \cite{Ral03} and information \cite{Enk01}. Realization of Schr\"odinger cat states of a few ions \cite{Mon96} and photons \cite{Bru96} has been awarded by the 2012 Nobel Prize. Cat states have also been realized as the superposition of few micro amperes of superconducting currents \cite{Fri00}. The main obstacle in the realization of large cat states is the extreme fragility of these states to loss, which inspires the application of fast operation within the ultra-stable medium. The fast operation has lead to the realization of a large cat state of 100 microwave photons in a waveguide cavity coupled to a superconducting qubit \cite{Vla13}. In the same spirit, the recent successes in dressing many-body systems to Rydberg level \cite{Zei16} that provides strong many-body interaction and the advances in the stability of Ultra cold atoms, open new windows for the generation of large spin cat states. 

The current paper proposes the generation of spin cat state within the Bose-Einstein condensate of ultra stable Rubidium atomic clock states, using the strong Rydberg interaction and the fast entanglement generation scheme using Lipkin-Meshkov-Glick type Hamiltonian \cite{Opa15, Mus15, Zoller,Vid11}. By detail optimization of all the parameters in  the scheme and by considering  destructive sources  namely  spontaneous emission, collective decoherence, level mixing, and simulating decoherence effects using Quantum Jump Monte Carlo, this proposal predicts fifty times  improvement in cat state size comparing to the state of the art realization of 14 spin cat states \cite{Mon11}.
Previous related, but distinct Rydberg based proposals  includes  Refs. \cite{Saf09,Opa12}, who performed detailed studies of the creation of moderate-size cat states using Rydberg blockade. The number of atoms is limited to of order ten in these schemes due to competing requirements for the presence and absence of blockade between different Rydberg transitions in the same ensemble. Ref. \cite{Muk11} briefly discussed the creation of moderate-size (15 atoms) GHZ type states in Strontium atom chains. The number of atoms in Ref. \cite{Muk11} is limited by unwanted transitions to other nearby many-body states \cite{Mukherjee-Thesis}. Ref \cite{Kha16} used the Kerr-type dressing interaction to make cat states in the order of 100 atoms. In the current paper, fast Lipkin-Meshkov-Glick type Hamiltonian in a Rydberg dressed BEC speeds up the entanglement generation process \cite{Zoller}  by a factor that is proportional to the size of cat state  comparing to the scheme of Ref. \cite{Kha16} and therefore results to the formation of 700 atoms cat state.

Coupling of Rydberg atoms and mechanical oscillators has been studied and used as a quantum control of mechanical systems \cite{Ant14,Yan15,Car14,Liu18,Meystre}.  The following part of this paper proposes to transfer the created spin cat state to  the  superposition of vibrational modes of two spatially separated mechanical oscillators using the dipole-dipole interaction between Rydberg atoms and charged cantilevers. 
Application of Rydberg-Cantilever coupling has been proposed in Ref.  \cite{Meystre} for making number superposition states of a cantilever, as well as entangled states of a pair of oscillators that are both limited in size to only two phonons by protocol design. The new proposal in this paper results to fifty times improvement in the size of entangled states that is limited by about 80\% Poisson probability of not loosing any phonon over the transition process. 
Thanks to the long range interaction, the long separation between the superposed elements would be ideal for the test of spontaneous wave-function collapse models \cite{Bas13}. In the same spirit other Rydberg assisted spatial superposition states of atoms has been proposed in \cite{Muk15,Mob13}.

The paper begins with the description of the scheme in Sec.~\ref{sec:Scheme} over which the bifurcation scheme and dressing potential are optimized in subsections \ref{sec:Bif} and \ref{sec:Dressing}. In Sec.~\ref{sec:decoherence} effects of de-coherences are quantified on the entanglement of the final cat state. Sec.~\ref{sec:size} gives an estimate on the achievable size of entanglement. Following that in Sec \ref{sec:error} destructive consequences of large atomic density i.e. level mixing and anomalous broadening are discussed and quantified. 
Finally in Sec. \ref{sec:mechanical} mapping the spin cat state to the mechanical mode is discussed.   
The paper is concluded by looking into the possible applications and avenues to the extension of the proposal.

\section{Scheme}
\label{sec:Scheme}

A bimodal  component Bose Einstein condensate of $N$ Rb atoms in clock states $|e\rangle$ and $|g\rangle$ \cite{Rie10,Li08} with populations $N_e$ and $N_g$ is considered. The mean field energy functional of the BEC $E[\psi_e,\psi_g;N_e,N_g]$ is given by ($\hbar \equiv1$)
\begin{equation}
\label{BEC}
E_{BEC}=\sum \limits_{i=e,g} N_i(\nu_i+\frac{N_i-1}{2} u_{ii}) + N_e N_g u_{eg},
\end{equation}
where $\nu_i=\int (-\frac{\bigtriangledown^2}{2m}+V_i)|\psi_i|^2$ is kinetic and trap potential energy of a single particle with wave-fuction $\psi_i$. Intra- and inter-component energies are given by  $u_{ij}=\int \frac{4\pi  a_{ij}}{m} |\psi_i|^2|\psi_j|^2$, for the scattering length of $a_{ij}$ and  atomic mass of $m$.

To enhance the nonlinear interaction, the $|e\rangle$ component of BEC with population $N_e$ is adiabatically dressed with the Rydberg level \cite{Hen13,Bal14,Hen10,Mau11,Hen12,Joh10,Gil14}.  
Rydberg dressing  of an ensemble accommodated well within the blockade radius, results to an effective light shift
that is a function of  enhanced Rabi frequency $\sqrt{N_{e}}\Omega_{r}$ \cite{kuzmich}.
 The effective light shift of the ground dressed state would be 
\begin{equation}
\frac{\Delta}{2}(1 - \sqrt{1+\frac{N_{e}\Omega_{r}^{2}}{\Delta^2}}),
\label{Eq:lightShift}
\end{equation}
 where $\Delta$ and $\Omega_{r}$ are the detuning and the Rabi frequency of the dressing laser. In the weak dressing regime $(w\equiv\frac{N_e\Omega_r^2}{\Delta^2}\ll 1)$ 
 one can Taylor expand the light shift in terms of $N_e$, 
\begin{equation}
\label{expansionLS}
E_{dressed}=\frac{\Delta}{2}[1 -  (1+\frac{1}{2} \frac{N_e \Omega_r^2}{\Delta^2} - \frac{1}{8} \frac{N_e^2 \Omega_r^4}{\Delta^4} +O(\frac{N_e \Omega_r^2}{\Delta^2})^3)].
\end{equation}
Within the fully symmetric subspace, the pseudospin of $J=N/2$ is considered. Considering the annihilation  operators $a_e$ and $a_g$ of the two hyperfine clock state modes $|e\rangle$ and $|g\rangle$, the Schwinger pseudo-spin $\hat{J}_z=\frac{\hat{a}_e^{\dagger}\hat{a}_e-\hat{a}_g^{\dagger}\hat{a}_g}{2}=(\hat{N}_e-\hat{N}_g)/2$  corresponds  to the population difference of the excited and ground state atoms. Two other  perpendicular basis  $\hat{J}_x=\frac{\hat{a}_e^{\dagger}\hat{a}_g+\hat{a}_g^{\dagger}\hat{a}_e}{2}$, $\hat{J}_y=\frac{\hat{a}_g^{\dagger}\hat{a}_e-\hat{a}_e^{\dagger}\hat{a}_g}{2i}$ are defined to fulfill the spin commutation relation $[\hat{J}_l,\hat{J}_m]=i\epsilon_{lmn}\hat{J}_n$.

\begin{figure} [t]
\raggedleft
\scalebox{0.34}{\includegraphics*[viewport=0 0 1050 500]{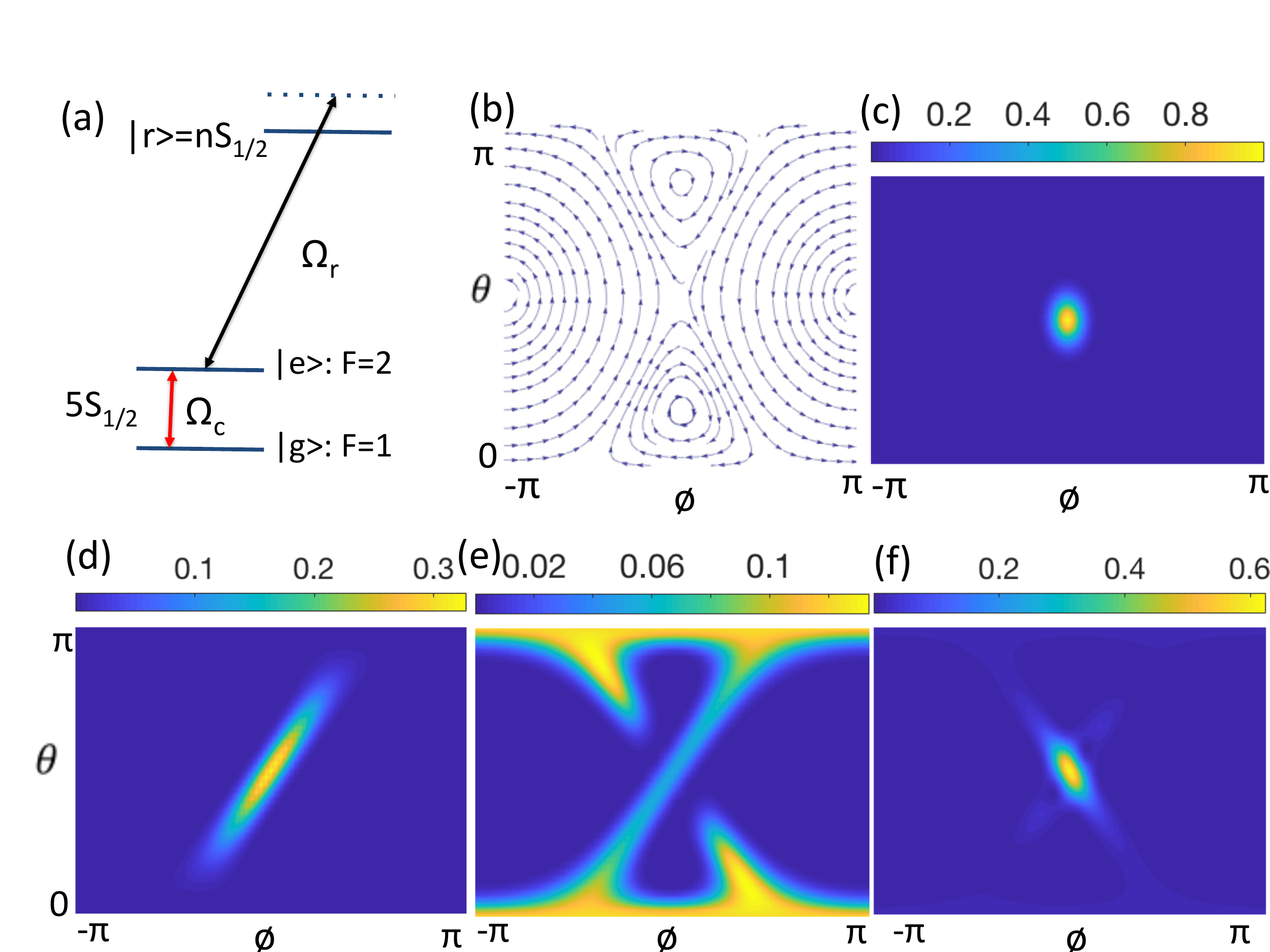}}
\caption{ (Color online) . Proposed scheme for creation of large spin superposition.
(a) Level scheme in Rubidium. The pseudo spin states are the
hyperfine levels in the ground state. An off-resonant laser field $\Omega_r$ dresses the excited state with the Rydberg level $|r\rangle$ via a ladder two excitation process. This creates a Kerr-type interaction between the
atoms in the excited state. (b) The vector field representation of Hamiltonian (Eq.~\ref{Hamiltonian}) under the simultaneous application of $\Omega_c$ and $\Omega_r$ and optimal condition $\Lambda=2$ and $\delta=0$, see Fig.~\ref{Optimized bifurcation}. The created bifurcation and bi-stabilities in (b) explains how an initial coherent spin state CSS (c), squeezes initially (d)  and eventually splits into two CSS pointing in opposite directions on the Bloch sphere (e). The final state could be approximated as the superposition of all the atoms being in the ground state and excited state. (f) Revival of the initial state could be seen after $2\tau_c$. The generated figure shows the case for 80 atoms.
} \label{Fig:Scheme}
\end{figure}



Considering a BEC that simultaneously is driven with both Rydberd dressing laser ($\Omega_r$, $\Delta$) and the other laser coupling the clock states ($\Omega_c$, $\Delta_c$) as shown in Fig. \ref{Fig:Scheme}; the effective Hamiltonian of the system in the  pseudo-spin notation would be of Lipkin-Meshkov-Glick type:
\begin{equation}\label{Hamiltonian}
H=\chi \hat{J}_{z}^{2}+\delta \hat{J}_{z}+\Omega_{c} \hat{J}_{x}.
\end{equation}
Non linear coefficient $\chi=(u_{ee}+u_{gg}-2u_{eg})/2+\chi_0(1-3\omega+15\omega^2/2)$ is approximately $\chi \approx \chi_0 \equiv \frac{\Omega^4}{16\Delta^3}$ in the regime of interest of this paper, see Sec. \ref{sec:size}. Linear rotation along $z$ axis with  $\delta=\nu_e-\nu_g+\Delta_c+(u_{ee}-u_{gg})(N-1)/2+\chi_0 N (-\omega^{-1}+1-9\omega^2+5\omega^4)$ could be made zero by adjusting $\Delta_c$, see Sec. \ref{sec:Dressing} for derivation.
 

Combination of linear rotation along the $x$ axis and a non-linear rotation along the $z$ axis, can lead to bifurcation and bistabilities that could be used in making cat states. Fig.~1b is the vector field representation of Hamiltonian (Eq.~\ref{Hamiltonian}) under the optimal condition
 $\Lambda=\frac{N \chi }{\Omega_{c}}=2$ and $\delta=0$ (see Sec. \ref{sec:Bif}). This figure 
shows how an initial coherent spin state (CSS) $| \Psi_{0} \rangle=\otimes_{i=1}^{N}\frac{(\left|g_{i}\right\rangle +\left|e_{i}\right\rangle )}{\sqrt{2}}=\left|J_{x}+\right\rangle$ (Fig. \ref{Fig:Scheme}c),  gets squeezed initially (Fig. \ref{Fig:Scheme}d)  and eventually splits into two CSS pointing in opposite directions on the Bloch sphere (Fig. \ref{Fig:Scheme}e). The final state could be approximated as the superposition of all the atoms being in the ground state and excited state $\left|GHZ\right\rangle =\frac{1}{\sqrt{2^{N}}}(\otimes_{i=1}^{N}\left|g_{i}\right\rangle +\otimes_{i=1}^{N}\left|e_{i}\right\rangle )$, see Sec. \ref{sec:Bif}.  

\subsection{Optimized Bifurcation Scheme}
\label{sec:Bif}
\begin{figure}
\raggedleft
\scalebox{0.46}{\includegraphics*[viewport=0 145 750 500]{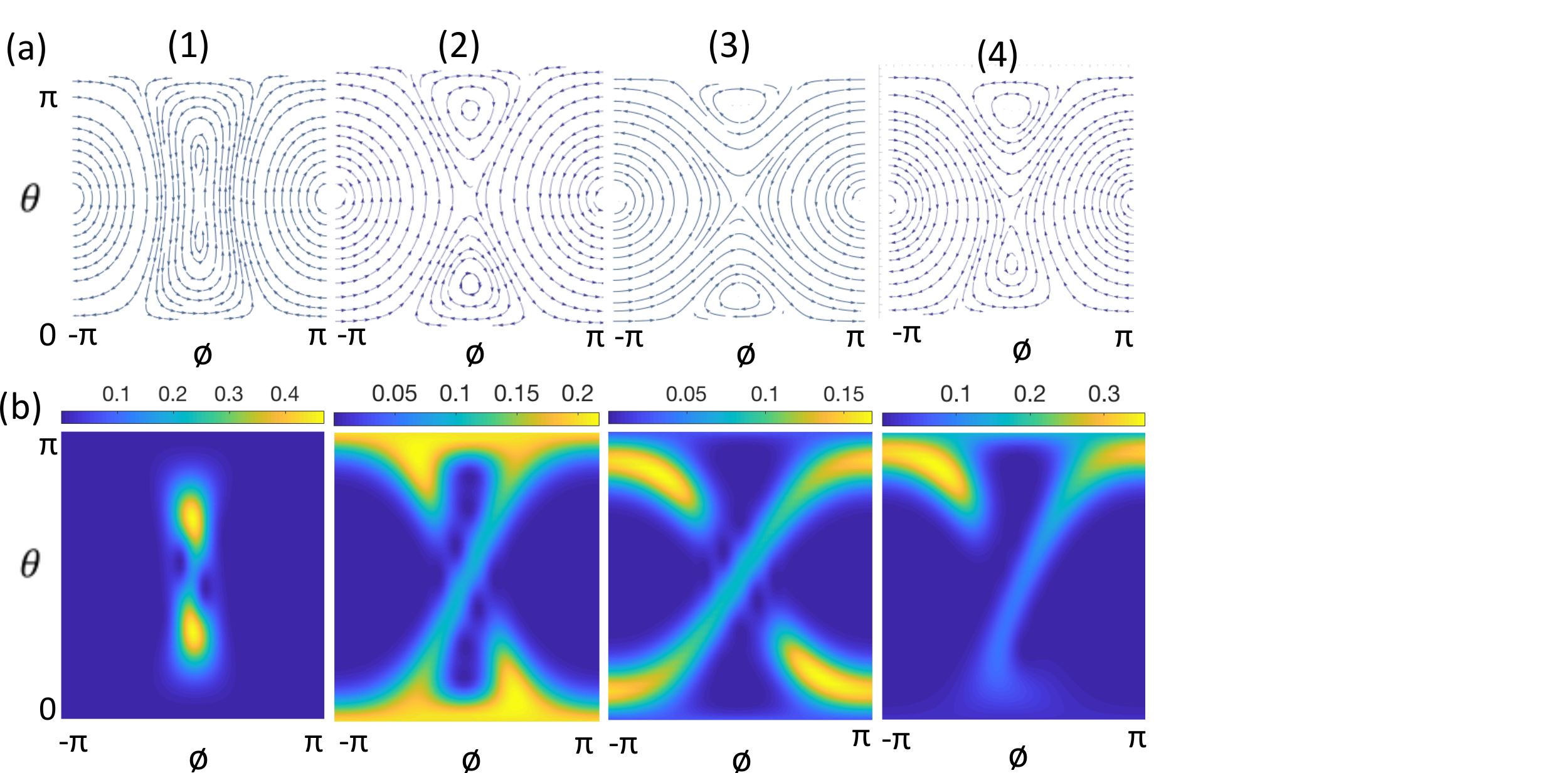}}
\caption{ (Color online) Optimized bifurcation parameters. (a) The trajectories  and  (b) Husimi Q-functions of 30 atoms  are plotted for different parameters $\Lambda=1.15, 2, 3$ in the columns (1-3).
Considering the cat state wave function of $|cat(\Delta \theta)\rangle=(|\frac{\pi-\Delta \theta}{2},\phi'\rangle+|\frac{\pi+\Delta \theta}{2},\phi'\rangle)/\sqrt2$, the angular separation in column (1-3) are $\Delta \theta=\{0.4\pi, 0.98\pi,  \, 0.8\pi \}$  and  $\phi'=\{ 0,\, 0, \, \pi\}$ with the fidelity of Fid=\{0.94, 0.42, 0.31\} and  Fisher information of  F/N=\{10.5, 20.3, 18\}. Odd orders of $J_z$ disturbs the symmetry in the two hemispheres. This fact can be seen in column (4) with $\Lambda=2$, $\delta'$=0.3.} \label{Optimized bifurcation}
\end{figure}

Semi-classical trajectories and Husimi Q-function of states evolved under Eq. \ref{Hamiltonian} are used for exploring the dynamics and optimization. 
In the limit of large ensemble, the quantum uncertainty of the CSS ($\propto N^{-0.5}$) can be neglected and the collective spin operators could be replaced by their expectation values $\hat{J}\rightarrow(\langle \hat{J}_x \rangle,\langle \hat{J}_y \rangle,\langle \hat{J}_z \rangle)=\frac{N}{2}\times (x,y,z)=\frac{N}{2} \times (\sin(\theta)\cos(\phi),\sin(\theta)\sin(\phi),\cos(\theta))$. 
Using the Ehrenfest theorem and factorization of operators' expectation values $\langle \{\hat{J}_x,\hat{J}_y\}\rangle =2\langle \hat{J}_x \rangle \langle \hat{J}_y \rangle$, the equations of motion in the spherical coordinate reads
\begin{equation}
\left(\begin{array}{c}
\dot{\theta}\\
\dot{\phi}
\end{array}\right)=\Omega_c\left(\begin{array}{c}
-\sin(\phi)\\
\Lambda \cos(\theta)-\cos(\phi)\cot(\theta)+\delta'
\end{array}\right),
\label{Fig:trajectories}
\end{equation} 
where $\Lambda=\frac{N\chi}{\Omega_c}$ and $\delta'=\frac{\delta}{\Omega_c}$.  The semi-classical explanation above are valid before the interference time at twice the cat creation time $2\tau_c$. Corresponding trajectories of Eq. \ref{Fig:trajectories} and Husimi Q-functions of 30 atoms evolved under Eq. \ref{Hamiltonian} are  plotted in Fig.~\ref{Optimized bifurcation}. The ideal cat state in this proposal with angular separation of $\Delta \theta_c$ could be written as $|cat\rangle=(|\frac{\pi-\Delta \theta_c}{2},\phi_c\rangle+|\frac{\pi+\Delta \theta_c}{2},\phi_c\rangle)/\sqrt2$ where the coherent spin state is defined as $|\theta,\phi \rangle=(1+|\eta|^{2})^{-j} \sum_{m=-j} ^{j} \eta^{j+m} \sqrt{\left(\begin{array}{c} 2j\\ j+m \end{array}\right)}|j,m\rangle$ \cite{CSS}, with $\eta=\tan(\theta/2)e^{-i\phi}$, and $|j,m \rangle$ being the Dicke state. To find the optimum dynamics the angular separation of cat states and their fidelities Fid=$|\langle \psi | cat\rangle|^2$ are compared in Fig. \ref{Optimized bifurcation}. For different dynamical parameter of $\Lambda=[1.15,\, 2, \, 3]$ the angular separations are  $\Delta \theta_c=\{0.4\pi, 0.98\pi,  \, 0.8\pi \}$  with  $\phi_c=\{ 0,\, 0, \, \pi\}$ and corresponding fidelities of  Fid=\{0.94, 0.42, 0.31\}. The Fisher information in columns (1-3)  of Fig. \ref{Optimized bifurcation} are  F/N=\{10.5, 20.3, 18\}, see below. 
These results show that $\Lambda=2$ results to $|GHZ \rangle$ state with maximum angular separation and entanglement and will be considered in the rest of the paper. 

\subsection{Measure of Macroscopicity} 
\label{sec:Fisher}
The macroscopicity of quantum state could be quantified using quantum Fisher information.  
Fisher information along the $J_z$, simplifies to the quantum variance of collective spin $F[\rho,J_{z}]=\langle J_z^2 \rangle - \langle J_z \rangle^2$.
Quantum limit of Fisher information $F>N$ for a linear collective transformation (here $J_z$) indicates the presence of entanglement \cite{Pez09}. 
The solid line in Fig. \ref{Fisher}a shows the rise of the Fisher information to $10Log_{10}(F/N)=22$dB over the cat generation time, close to the ultimate Heisenberg limit of 24.7dB for $N=300$ atoms. 

\subsection{Optimal Dressing Strength}
\label{sec:Dressing}
Optimization of the dressing strength has an important role in the success of the scheme. While stronger dressing results to a stronger interaction and positively reduces the process time, it increases the effect of higher orders of non-linearities in Eq. \ref{expansionLS}.
The  Hamiltonian corresponding to Eq. \ref{expansionLS} in the spin operator bases $\hat{N}_e=\hat{J}_z+N/2$ would read
\begin{eqnarray}
\label{expansionHamiltonian}
&&\hat{V}_{in}=\chi_0[N(-w^{-1}+1-9w^2+5w^4)\hat{J}_z\\ \nonumber
&&+(1-3w+\frac{15w^2}{2})\hat{J}^2_z+\frac{2}{N}(-w^2+5w^4)\hat{J}^3_z+O(\hat{J}^4_z)]
\end{eqnarray}
where $\chi_0=\frac{\Omega^4}{16\Delta^3}$. 
The odd orders of $J_z$ generates counter rotation in the south and north hemispheres, resulting in an asymmetry in the direction of flow and unequal superposition of CSS in the south and north poles, see Fig.~\ref{Optimized bifurcation}. The effect of first order $\delta \hat{J}_z$ could be compensated by adjusting  $\Delta_c$, while higher odd orders would be negligible in the weak dressing regime, see Eq.~\ref{expansionHamiltonian}. Figure \ref{Fisher}b, quantifies the Fisher information of the generated cat state as a function of dressing strength for a wide range of cat sizes. The scaling of the optimum dressing strength $w$ as a function of the cat size is represented in Fig.~\ref{Fisher}e.

\section{Decoherence}
\label{sec:decoherence}

Cat states are very fragile to de-coherence. 
While one atom loss completely destroys the coherence of cat state, the probability of one atom loss over the cat generation process is not necessarily fatal since at the early stages, the fragile superposition has not been formed and the state is more robust against de-coherence.

\begin{figure} [t]
\raggedleft
\scalebox{0.47}{\includegraphics*[viewport=0 0 800 560]{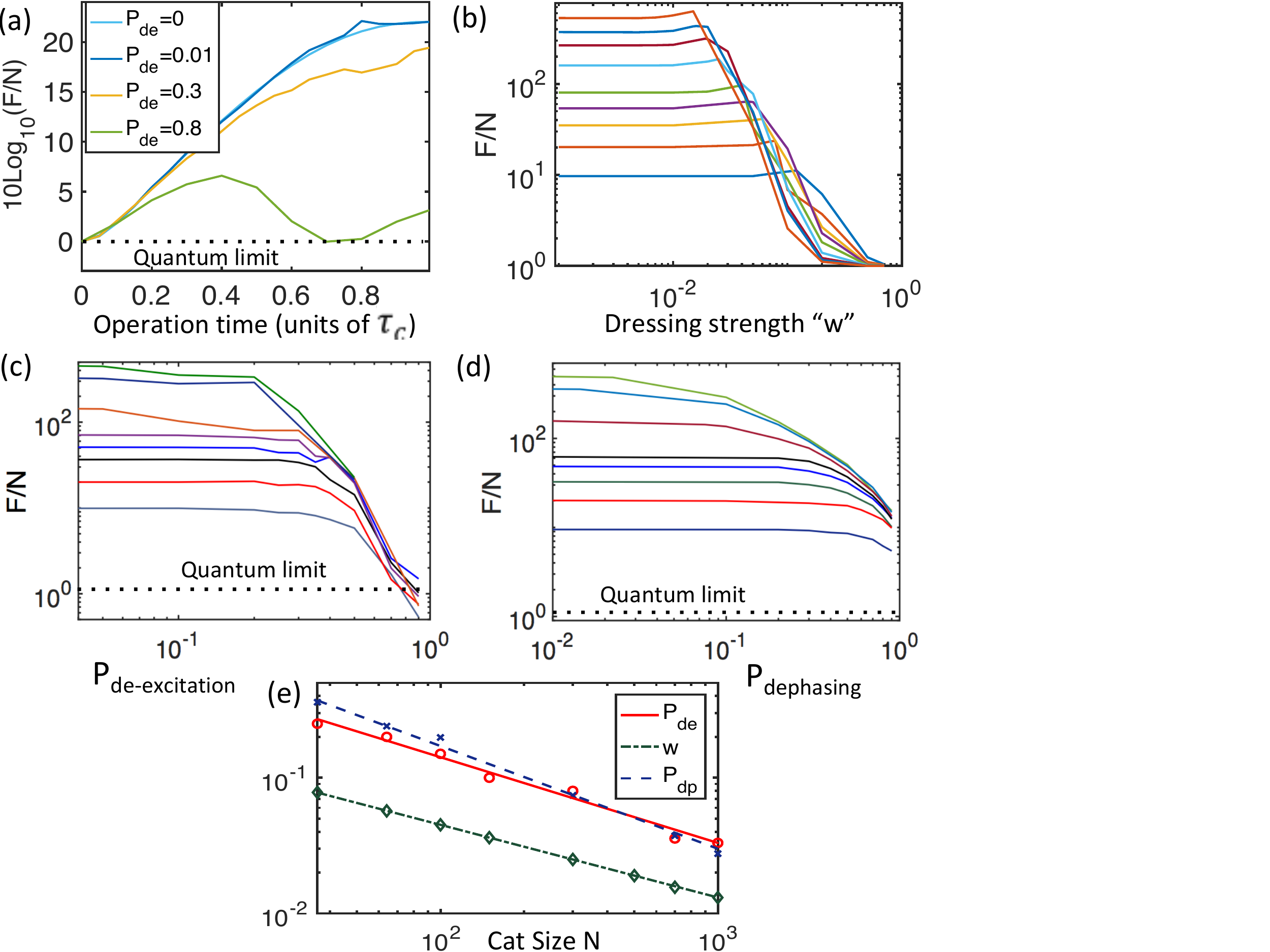}}
\caption{ (Color online) Effects of dressing strength and de-coherence on the entanglement generation. a) The evolution of Fisher information over cat state generation of N=300 atoms, for different  de-excitation probabilities of $P_{de}(\tau_c)=\{0, 0.01, 0.3, 0.8\}$. Fisher information of the generated cat states is plotted as a function of (b) dressing strength and also Poisson probabilities of (c) de-excitation (d) and de-phasing. The lines from bottom to top are corresponding to the ensemble sizes of $N=\{ 16, 36, 64, 100, 144, 300, 700, 1000\}$ atoms. 
The Monte-Carlo simulations have been averaged over 160  separate runs. (e) The trend of optimal dressing strength  $w$ as well as de-excitation and de-phasing rates that reduce the Fisher information by 10\% are plotted as a function of cat size.} \label{Fisher}
\end{figure}

\begin{figure}
\raggedleft
\scalebox{0.45}{\includegraphics*[viewport=90 125 740 460]{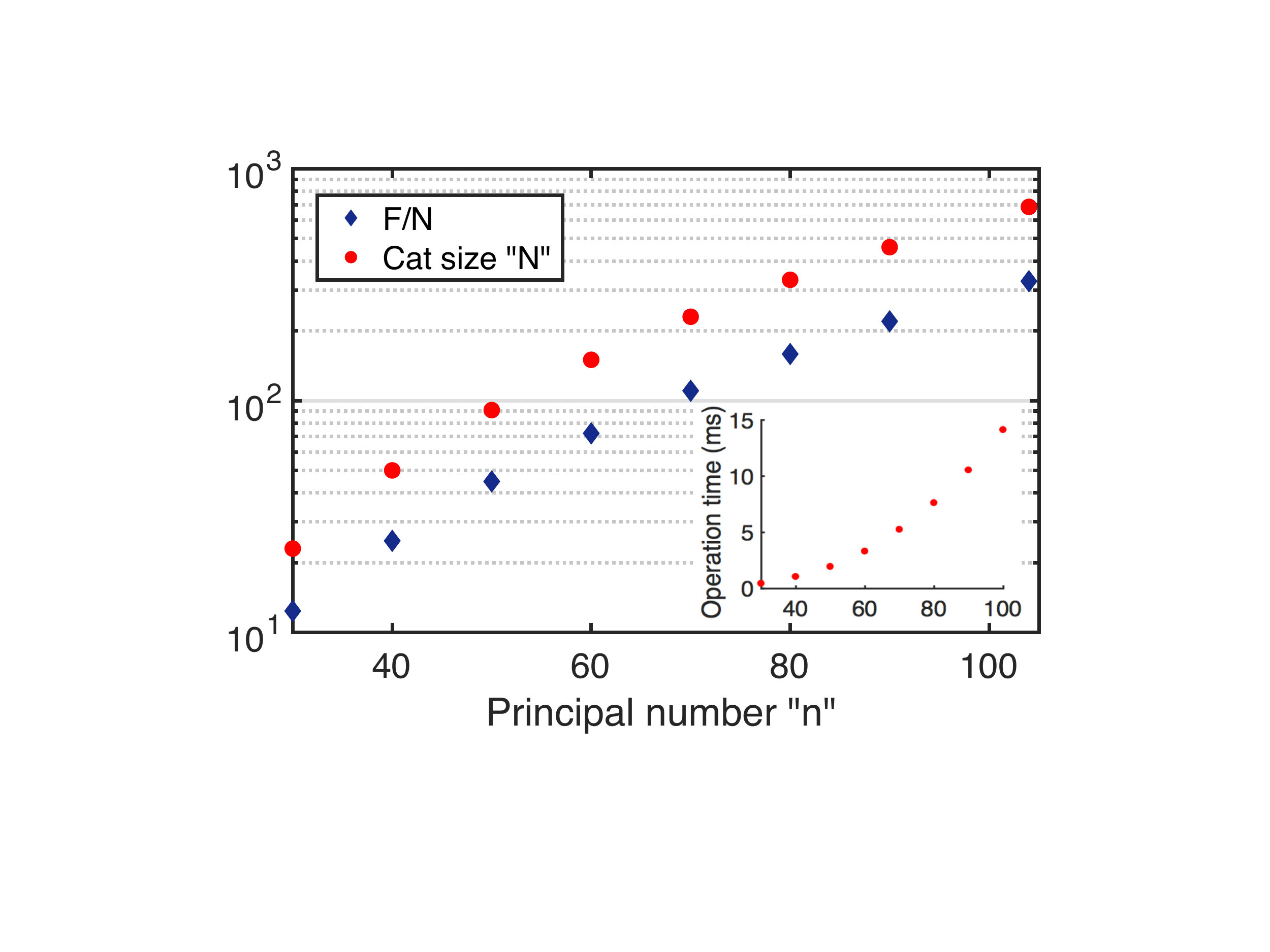}}
\caption{ (Color online) Maximum achievable cat size and corresponding Fisher information as a function of the principal number $n$ of the Rydberg state. 
The de-excitation and de-phasing probabilities and dressing strength are set according to Fig.~\ref{Fisher}e 
Blockade radius is set to be three times larger than the trap diameter to ensure the homogeneity of interaction \cite{Kha16}. 
The inset shows the required cat creation time.
Without cryogenic environment achievable cat size would reduce from $700$ to $550$ atoms.}\label{Nvsn} 
\end{figure}

The main sources of de-coherence come from the Rydberg level. The de-coherences of the intermediate levels used for  $|e \rangle -|g \rangle$ coupling and dressing could be made negligible due to the weak Rabi frequency requirement. 
Dressing the excited state, slightly admixes the Rydberg level resulting in $(|\tilde{e} \rangle \approx |e\rangle+ \frac{\Omega_r}{\Delta } |r\rangle)$. 
Under weak dressing the effective de-coherence rate of the system would be $w \gamma_r $, where $w=\frac{N \Omega_r^2}{2 \Delta ^2}\ll 1$ is the expected Rydberg population in the ensemble ($N_e\approx N/2$ due to the symmetry in two sides of equator). Decay rate of the $nS_{1/2}$ Rydberg state of $Rb$ is given by $\frac{\gamma_r}{2 \pi}=\frac{116}{n^{*3}}$ MHz \cite{Bet09}. Here application of cryogenic environment
\cite{Cryogenic} is assumed to suppress the effects of blackbody radiation (BBR) on Rydberg decoherence and their effects are discussed in Fig. \ref{Nvsn}. 
Entire $1/e$ lifetime of $^{87}Rb$ BEC  could be as long as 3 min \cite{Sch04}. In the other word taking into account the one, two and three body scattering, the  Poisonian probabilty of not lossing any atom over the cat creation process is in the range of 94\%-99.99\% over the period of interest in Fig. \ref{Nvsn} which is negligible. 


The spontaneous emissions from the Rydberg level $|r \rangle$ and  the intermediate levels  will either directly or indirectly end up to one of  $|e\rangle$ or $|g\rangle$ splittings of the ground state. In the absence of external field the mentioned decoherences split with equal rates into a de-phasing $\hat{c}_{dp}=|\tilde e\rangle \langle \tilde e|$ and de-excitation $\hat{c}_{de}=|g\rangle \langle \tilde e|$ Lindblad terms in the dressed level bases $ |\tilde e\rangle =| e\rangle+\frac {\Omega}{\Delta}|r\rangle$, where  in the weak dressing regime $|\tilde e\rangle \approx | e\rangle$. Effects  of the mentioned decoherences  on the Fisher information is studied using the quantum jump Monte-Carlo simulation \cite{Monte-Carlo}
of the master equation $\dot{\rho}=-i[H,\rho]+\sum\limits_{i=de,dp}\gamma_i (\hat{c}_i \rho \hat{c}_i^{\dagger}-\frac{1}{2} \{\hat{c}_i\hat{c}_i^{\dagger},\rho\})$ where $H$ is defined in equation \ref{Hamiltonian}.
 Figure~\ref{Fisher}c(d) plots the Fisher information at the cat generation time $\tau_c$ as a function of de-excitation (de-phasing)   probability $P_{de(dp)}(\tau_c)=1-P_0(\tau_c)$, where $P_0=e^{-w \gamma_{de(dp)} \tau_c}$ is the Poisson probability of not de-exciting (de-phasing) any atom over the process. 
Fig.~\ref{Fisher}e shows the scaling of $P_{de(dp)}$ verses cat size $N$ that results to 10\% reduction of Fisher information.   
\section{Estimate of the Size of Entanglement} 
\label{sec:size}
Taking into account the mentioned imperfections, Fig.~\ref{Nvsn} shows the achievable cat size as a function of the principal number $n$. Up to around $n \sim 100$, the size increases with $n$.
Higher $n$ leads to a stronger interaction, hence allowing weaker dressing, and to a smaller loss, favouring the creation of larger cats. However, for $n \sim 100$ the decreasing separation of the neighboring Rydberg levels (which scales like $n^{-3}$) limits the detuning and hence the interaction strength. This will prevent  the generation of larger cat states at higher principal numbers. 
In Fig.~\ref{Nvsn} dressing strength and de-coherence terms are set by the scales represented in Fig.~\ref{Fisher}e. For the range of dressed Rydberg levels $n=30 \rightarrow100$ in Fig.~\ref{Nvsn}, dressing Rabi frequency and detuning would change over $\Omega_r=0.3MHz \rightarrow 10MHz$,  $|\Delta|=2MHz \rightarrow 650MHz$ and the non-linear coefficient would evolve as $\chi_0=6kHz \rightarrow 23Hz$.
Dressing to an $S$ orbital is desired due to its isotropic interaction in the presence of trap fields. 

\section{Possible Sources of Error}
\label{sec:error}
High density of BEC ensembles raises concerns about level mixing and anomalous broadening at small inter atomic separations.

\subsection{Level Mixing}
\begin{figure}
\raggedleft
\scalebox{0.60}{\includegraphics*[viewport=120 40 900 482]{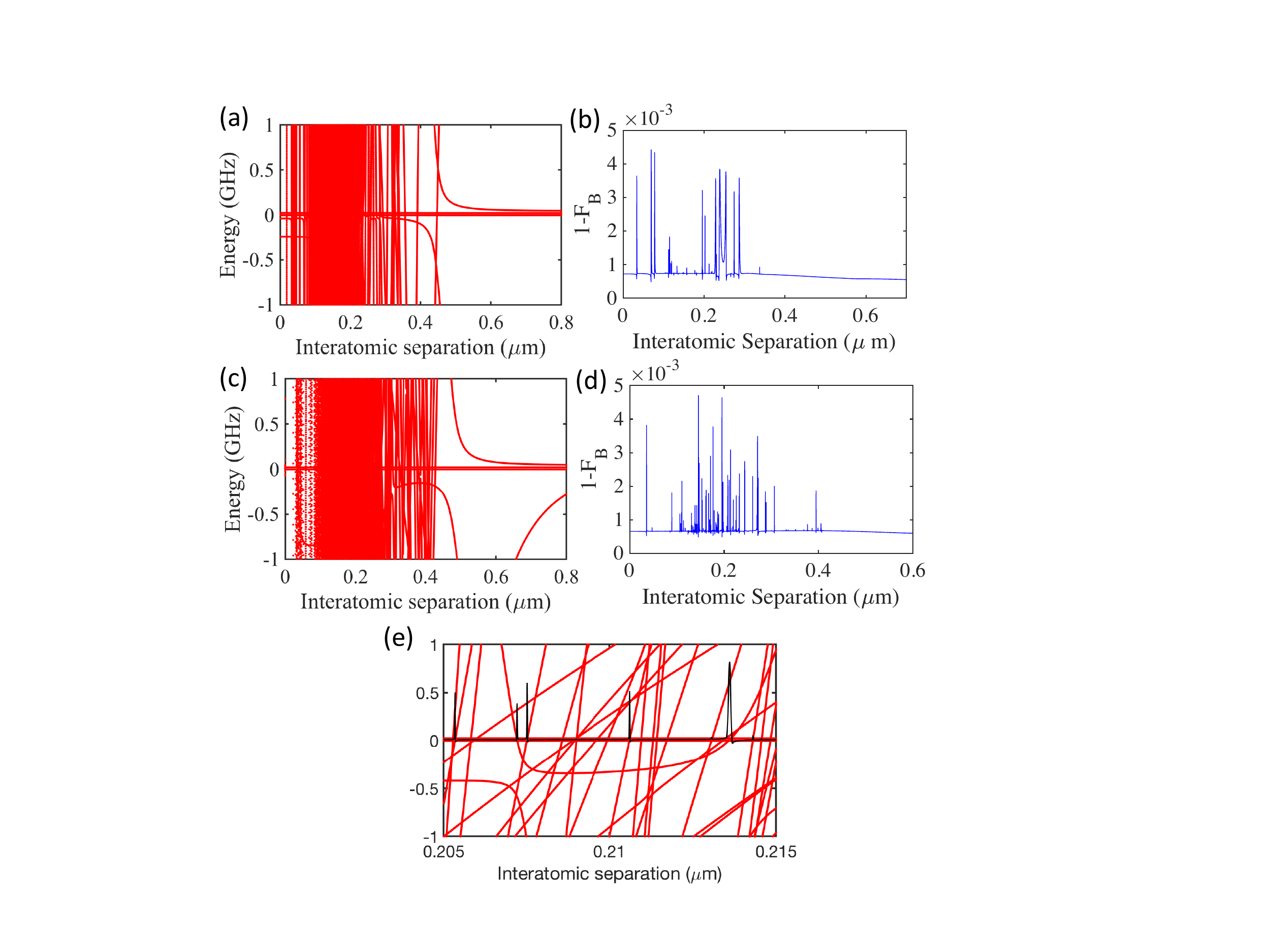}}
\caption{ (Color online) Level mixing and Blockade in-fidelity.
 (Left column) Eigen energies of dressing Hamiltonian (Eq. \ref{Eq:dressing})  close to the ground state dressing energy  (Eq. \ref{Eq:lightShift}) over close interatomic separation. First  and second rows are  corresponding to  interatomic orientations of $\theta=\{0, \, 90\}$ relative to quantization axis. 
(Right column) Represents the perseverance of blockade, mainly due to the dilute intensity of the optically accessible pair states.
 Figure (e) is a zoomed in version of (c,d) where the black line is the map of infidelity on the Eigen energy red (light gray) lines to match the position of minor infidelity peaks with the avoided crossing points. The graph is for dressing to $|50S_{1/2}1/2\rangle$ with $\Omega_r=3.3$MHz and $\Delta=-23$MHz
} \label{FIGLevel_mixing}
\end{figure}

The strong level mixing at short distances has raised questions about the perseverance of blockade at the avoided crossing points \cite{LevelMixing,LevelMixing2}. Fig.~\ref{FIGLevel_mixing} is a sample calculation representing the evolution of blockade efficiency over close inter-atomic separation featuring strong level mixing. By considering thousands of coupled Rydberg pairs  and comparing two cases evolved under Hamiltonians with and without level mixing, reported results are aligned with the observation of blockade perseverance at the intense level mixing region \cite{LevelMixing3}. These results suggest that while the interaction could bring some of the optically accessible Rydberg pairs into the resonance with the dressing laser, diluted density of the corresponding pairs makes the blockade infidelity benign.

Here is a more detailed discussion of the performed calculation.   Dressing laser couples the clock state $|e\rangle$, with the optically accessible Rydberg levels $|r^i\rangle$ with $\Omega_i$ Rabi frequencies  and $\Delta_i$ detunings. In the pair bases a system of two atoms evolves under the effective dressing Hamiltonian 
\begin{equation}
H_{dress}=H_{s}+H_{d}+H_c
\label{Eq:dressing}
\end{equation}
where
\begin{eqnarray}
&&H_s=\sum_i [\sqrt{2}\Omega_i (|ee\rangle \langle r^i e_{+}|+h.c.)+\Delta_i |r^i e_{+}\rangle \langle r^i e_{+}| ] \\ \nonumber
&&H_d=\sum_{i,j}[\sqrt{2}\Omega_j (|r^i e_{+}\rangle \langle r^i r^j|+h.c.)+(\Delta_i+\Delta_j)|r^i r^j\rangle \langle r^i r^j| ] \\ \nonumber
&&H_c=\sum_{i,j,k,l}\frac{C_3^{ij,kl}(\theta)}{R^3}(|r^l r^k\rangle \langle r^i r^j|+h.c.)
\end{eqnarray}
 are single, double excitation  and pair coupling Hamiltonians. Summation goes over all possible Rydberg states and $ |r^i e_{+}\rangle=(|r^i e\rangle+|er^i \rangle)/\sqrt2$ denotes symmetric two particle states. Dipole-dipole coupling constant $C_3^{ij,kl}(\theta)$ between different pairs is a function of  the relative orientation $\theta$ of the interatomic separation and the quantization axis defined by the dressing laser. 
A pair of close atoms that are dressed  to $|nS,nS \rangle$ Rydberg state, could get coupled to other dipole allowed pair states that are close in energy under the coupling Hamiltonian $H_c$. The product pairs would couple to their neighbors' and this would continue to couple a huge number of pairs. 
The first column of Fig.~\ref{FIGLevel_mixing} shows the eigen energies of Eq. \ref{Eq:dressing} close to the ground state dressing energy, see Eq. \ref{Eq:lightShift}. The number of coupled pairs are intense resulting to the formation of spaghetti of levels at the neighboring lattice cites.

To truncate the huge number of pairs, only those with strong interactions comparable to pair energy separations are considered in Fig.~\ref{FIGLevel_mixing} for the case of dressing to $n=50$. This works since transition matrix elements decreases quickly with $\Delta n$. At $n=50$ significant level crossing happens at the separation region of interest, and going to higher principal numbers would only make the calculation more intense and therefore not presented in here. Two rows in Fig.~\ref{FIGLevel_mixing}  are  corresponding to sample interatomic orientations of $\theta=\{0,  \, 90\}$ relative to quantization axis,  where the number of considered coupled pairs are $\{1000,  \, 3000\}$ reflecting the pair selection rules of $\Delta M=\{0; \, 0,\pm 2 \}$ for the respective angles ($M=m_{1}+m_{2}$ is the secondary total angular momentum number of a Rydberg pair).
Evolution of blockade efficiency over the desired interatomic separation  is plotted in the right column of Fig.~\ref{FIGLevel_mixing}, calculated by comparing the states evolved under the dressing Hamiltonian with level mixing in Eq. \ref{Eq:dressing} and the  ideal case with perfect blockade i.e. $H_{dress}=H_s$. 
 As one can see there is a minor in-fidelity at the avoided crossing points.
This is mainly due to the fact that a dilute portion of the coupled pair states might be optically accessible by the dressing laser $(H_s, H_d)$ and the strong interaction $(H_c)$ could bring them in resonance with the field. 
One should note that at short distances around LeRoy radius, assumed dipole-dipole interaction is only a toy model that represents a strong interaction. Including higher orders of interaction expansion i.e. quadrupole-quadrupole and so on, extend the selection rules, making the level mixing significantly more intense and the ratio of the accessible pairs gets more diluted; therefore blockade is still expected to be preserved. 
 

\subsection{Collective Line-Broadening}
 High density could also lead to linewidth broadening due to the strong dipole-dipole interaction between the target Rydberg levels and BBR-induced populations in neighboring levels \cite{Anomalous broadening}. However, weak dressing and relatively small atom number in our scheme make the probability of populating neighboring levels very small. For example for the environment temperatures of 3K \cite{Cryogenic}, 300K and for the corresponding achivable cat sizes of 700 and 550 at $n=104$ the probability of not populating the strongly interacting neighboring Rydberg levels over the process is $P_{BBR}(0)=exp(-w\gamma_{_{BBR}}\tau_c)=$99.95\% and 97.21\% respectively. It could be concluded from the experimental measurements of \cite{Zei16} that for $P_{BBR}(0)> 0.82\%$, the anomalous broadening effect is benign.  

\section{Mechanical Cat State}

\begin{figure}
\raggedleft
\scalebox{0.34}{\includegraphics*[viewport=100 10 900 460]{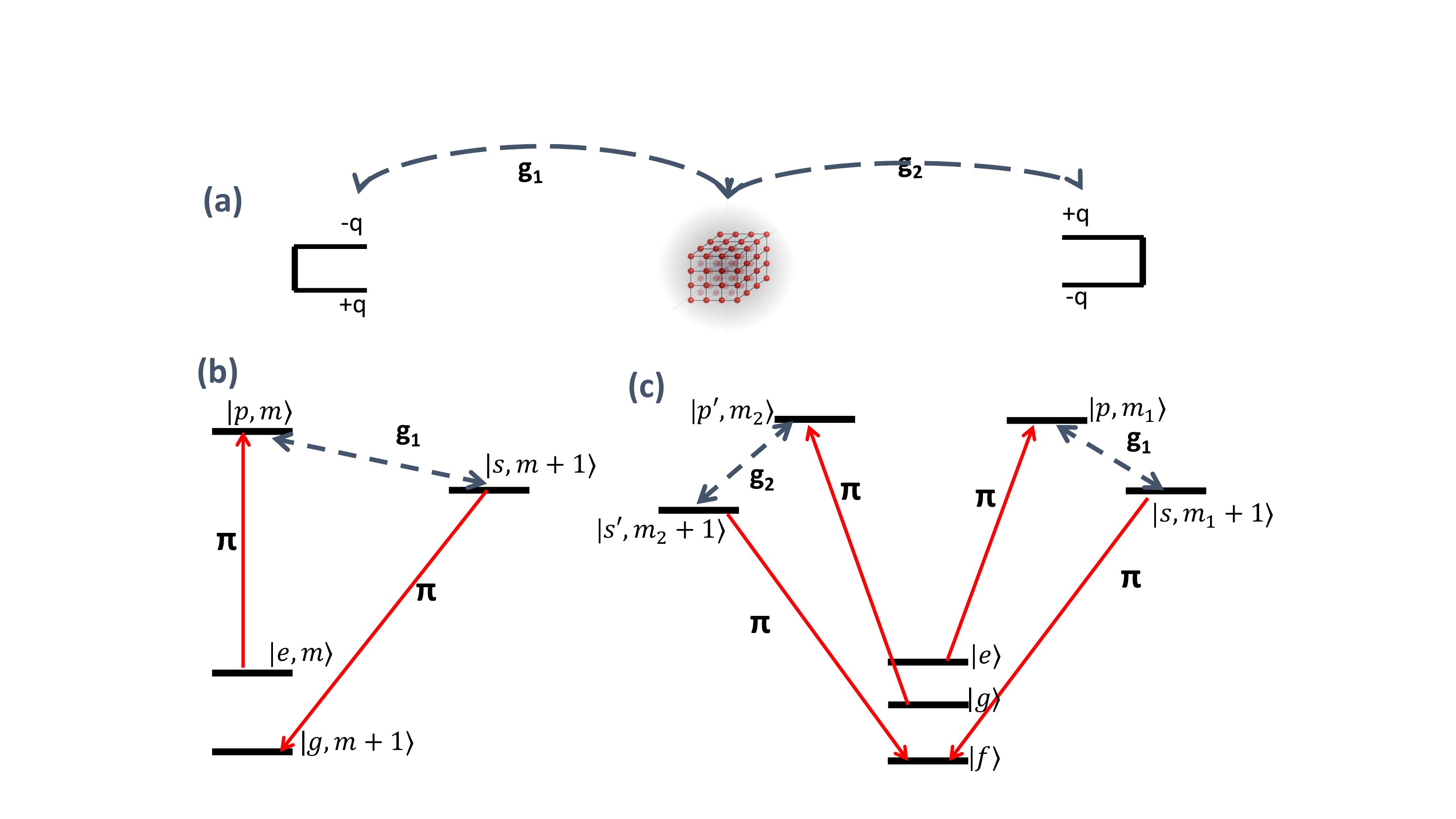}}
\caption{ (Color online) Proposed scheme for mapping the spin cat state to the mechanical system. (a) Setup includes two electrically-charged bridge-cantilevers coupled  with the Rydberg excited superatom via dipole-dipole interaction. (b,c) Level scheme. Rubidium clock states are labeled by $|e \rangle$ and $|g\rangle$. Collective Rydberg excitation with S and P orbitals are labled by $|s \rangle$ and $|p\rangle$. (b) Accommodating the ensemble within the blockade radius, applying the first $\pi$ pulse creates single Rydberg excitation in a collective form (superatom). The existing superatom provokes a single phonon in resonant mechanical oscillator over $\pi/g$ atom-cantilever coupling time. Following that a $\pi$ pulse transfers the Rydberg atom to the ground state. Consequent repetition of this cycle transfers the initial spin cat state to the mechanical system, resulting to the superposition of the presence and absence of N phonons on one cantilever. 
(c) The population of $|e \rangle$ and $|g\rangle$ would transfer to a third long lived state via different intermediate Rydberg levels that are in resonance with different cantilevers. Consequent repetition of this process results to the superposition of  $N$ phonons being in the right or left cantilever. } \label{Optomechanics}
\end{figure}

\subsection{Scheme}
Finally the generated cat state could be mapped  to the mechanical system.
Dipole-dipole coupling between Rydberg atom and charged cantilevers has been used for creation of phononic quantum states \cite{Meystre}. Here, this proposal is based on the application of atom-resonator coupling, for the transfer of already existing spin cat state  to the mechanical system. The scheme as demonstrated in Fig.~\ref{Optomechanics}a consists of two cantilevers equally far separated from atomic ensemble. The edges of the bridge like cantilevers are electrically charged to create dipoles. Cantilever-trap separation is much larger than the size of elements.

Interaction between a Rydberg atom and cantilever can be explained by the Hamiltonian of
$H_{C}+H_{A}+V_{A-C}$
where 
$H_{C}=\sum_{l=1,2}  \omega_l \hat{b}_l^{\dagger}\hat{b}_l,$ 
explains the free evolution of the fundamental mode of motion of the right $l=1$ and left  $l=2$ cantilevers with frequency $\omega_l$ and $\hat{b}$ is the  phononic annihilation operator. Atomic free evolution is given by  $H_{A}= \omega_s \hat{\sigma}_{ss}+\omega_p \hat{\sigma}_{pp}$ where  $\omega_s$ and $\omega_p $ are the energies of  $|s \rangle$ and $|p \rangle$ Rydberg states and $\hat{\sigma}_{pp}=|p\rangle \langle p|$ is the projection operator. Dipole-dipole  coupling between   $l$th cantilever and Rydberg atom is given by 
 \begin{equation} 
V_{A-C_l}=\frac{1}{4\pi\epsilon_0}(\frac{\vec{\mu}_{at}.\vec{\mu}_{l}}{R^3}-3 \frac{(\vec{\mu}_{at}.\vec{R})(\vec{\mu}_{l}.\vec{R})}{R^5})
 \label{dipoleCoupling}
\end{equation}
where $\vec{\mu}_{at}$ and $\vec{\mu}_{l}$ are atomic and cantilever dipole moments and $\vec{R}$ is the atom-cantilever separation. Here the atomic quantization axis and cantilevers' plate separation $\vec{d}$ are considered along the $z$ axis perpendicular to atom-cantilever separation ($\vec{R}$) leaving only the first term in Eq. \ref{dipoleCoupling}. The cantilever dipole moment is defined  as 
$\mu_{l}(t)=Q (d_l+z_l(t))$.
The constant dipole term, caused by original plate separation  ${d}$ cancels with the opposite charges of the second cantilever leaving the vibrational contribution along the $z$ axis. The remaining dipole term caused by the zero point motion of the resonator mode is given by $\mu_{l}=Q (\sqrt{\frac{1}{2m_l \omega_l}})$ \cite{Meystre}
where $Q$ is cantilevers' charge and $m_l$ is resonators' effective mass.  Atomic dipole transition along the $z$ axis  can be calculated as $\mu_{at_z}=e\langle n', l', j', m'_j  | z| n, l, j, m_j \rangle=(-1)^{j+l'-1/2}C_{jm10}^{j'm'}\sqrt{2j+1}\left\{ \begin{array}{ccc}
l & 1/2 & j\\
j' & 1 & l'
\end{array}\right\} \langle n', l' || r|| n, l\rangle
$  \cite{Saffman}
where $C_{j_1m_1j_2m_2}^{JM}$ are Clebsch-Gordan coefficients and curly bracket is the Wigner-6j symbol. The reduced matrix element is 
$\langle n', l' || r|| n, l\rangle=\sqrt{2l+1}C_{l010}^{l'0} \int r R_n'l'(r)R_nl(r)dr$ where $R_nl(r)$ is the radial wave-function that is numerically calculated by the Numerov method. 
Tuning the cantilever's frequency in resonance with a specific Rydberg dipole transition $\omega_l=\omega_s-\omega_p$, and within the rotating wave approximation, atom-cantilever coupling's Hamiltonian simplifies to 
\begin{equation}
V_{A-C_l}= g_l (\hat{b} \hat{\sigma}_{ps}+\hat{b}^{\dagger} \hat{\sigma}_{sp}),
\end{equation}
where $g_l=\frac{\mu_{at}\mu_{l}}{4\pi\epsilon_0R^3}$ is the coupling strength and $\hat{\sigma}_{ps}=|p \rangle\langle s|$ is the transition operator. 

\label{sec:mechanical}

After cooling the cantilevers to their ground state $|m=0 \rangle$, one can use the scheme presented in Fig.~\ref{Optomechanics}b to map the spin cat state  $(|e=N\rangle+|g=N\rangle)/\sqrt{2}$ to the mechanical system. The steps are as following:
1) Applying a fast $\pi$ pulse in resonance with $|e\rangle \leftrightarrow |p\rangle$ transition, creates a single collective Rydberg excitation, since the entire medium is within the blockade radius, resulting to $(|e=N-1,p,m=0\rangle+|g=N,m=0\rangle)/\sqrt{2}$.
2) Afterwards atom cantilever coupling will create a phonon in the mechanical mode, transferring the state to $(|e=N-1,s,m=1\rangle+|g=N,m=0\rangle)/\sqrt{2}$ within $\tau=\frac{\pi}{g}$ interaction time.
3) Following that a strong $\pi$ pulse transfers the Rydberg  $|s\rangle$ state to the ground state $|g\rangle$ resulting in $(|e=N-1,g=1,m=1\rangle+|g=N,m=0\rangle)/\sqrt{2}$. 
The blockade effect makes the  collective Rabi oscillation between the ground state and  single collective Rydberg excitation $|s\rangle$ in step (3). As a result, presence of $|s\rangle$ state at the begining of step (3) makes the $\pi$ pulse to only de-excite the $|s\rangle$ state and do not act in the opposite direction.
Repeating the mentioned cycle  in Fig.~\ref{Optomechanics}b for N times transfers the spin cat state to the mechanical system resulting in $|g=N\rangle(|m=N\rangle+|m=0\rangle)/\sqrt{2}$.

Fig.~\ref{Optomechanics}c is a scheme that maps the spin cat state to the mechanical mode of two spatially separated cantilevers. 
The population of $|e \rangle$ and $|g\rangle$ would transfer to a third long lived state  via different intermediate Rydberg levels that are in resonance with different cantilevers. Consequent repetition of this process results to the superposition of  $N$ phonons being in the right or left cantilevers $(|m_1=N\rangle+|m_2=N\rangle)/\sqrt{2}$.

A possible example for the realization of these schemes is the coupling between Rydberg transition of $179 S_{1/2} 1/2-179 P_{3/2}1/2$ in $Rb$ atoms with the transition frequency $590MHz$, dipole moment $\mu_{at_z}=15620 e a_0$ and a diamond bridge cantilever with the dimensions $(0.5,0.05,0.05) \mu m$, density of $\rho=3\times10^{-3} kg cm^{-3}$, Young modulus of $E=1050Gpa$ \cite{cantilever-fabrication} with the fundamental frequency of $590MHz$ \cite{M-frequency} and dipole moment of $\mu_l=1.7\, 10^{-3}Qa_o$  \cite{Meystre}. When cantilevers are $5\mu m$ separated from the ensemble and their plates are charged by $q=\pm 7 \times 10^{3} e$, coupling constant between the explained single Rydberg atom and cantilever in its ground state will be $g/2\pi=1MHz$. 

\subsection{Size Estimate of the Entangled State and  Decoherence Discussion}

Since the entire atomic ensemble is accommodated within the blockade radius, the collective nature of single excitation representing the superatom leads to a collective enhanced coupling of $\sqrt{N_{e}}$. Effective resonator-Rydberg superatom coupling between initial $|p,m\rangle$  and final $|s,m+1\rangle$  states is also enhanced by the number of phonons \cite{Meystre} to $g \sqrt{N_{e}} \sqrt{m+1}$. Therefor the required time for transferring the cat state is given by $\tau = \sum \limits_{m=0}^{N-1} \frac{\pi}{g \sqrt{N-m} \sqrt{m+1}}$ for Fig. 6b and 2$\tau$ for Fig. 6c since the two cycle could not be run simultaneously within the blockade radius. While the coupling of superatom is enhanced by $\sqrt{N_{e}}$, the spontaneous emission from the Rydberg level scales as a single particle decay rate \cite{Zoller-optomechanics}.  

The dominant source of decoherence in the cat state transition process is mechanical loss. Fabrications with high quality factors of $Q=6 \times 10^{6}$ \cite{Quality-factor}, corresponds to $\Gamma_{m}/2 \pi=96Hz$. The heating rate of cantilever is $\Gamma_{m,T}/2 \pi=300Hz$ at the temperature of $T=90mk$.  
The other source of loss is the spontaneous decay of Rydberg level which is $\Gamma_{R}/2 \pi=10 kHz$. Although Rydberg decay rate is greater than mechanical decoherence rate, mechanical loss is the major decoherence channel because we only have the maximum of single population in the Rydberg level over the transferring process but multi population in the phononic state. Using the setup in Fig.~4b and 4c with a coupling constant of $g/2\pi=1MHz$, one can transfer a spin cat state with $100$  atoms to a mechanical cat state in $1.3\mu s$ and  $2.6\mu s$ over which the Poisson probability of not losing any qubit is 85\% and 70\% respectively.

\quad

\section{Outlook}
The extension of this work to the solid-state environment could improve  the size of the entangled state. Rydberg dressing of $1$S excitons in $Cu_2O$, to the $nP$ Rydberg levels via $CO_2$ lasers results to large soft cores \cite{Bayer,Kha17}. Considering the $1$A Bohr radius size of $1S$ excitons that are confined to the copper ions and $10\mu m$ Blockade radius of $24P$ Rydberg state, the number of entangled excitons could improve upon the future progresses in reducing the background phononic noise of Rydberg spectrum.
This proposal could also be applied in the clock state of alkaline earth atoms resulting to huge energy cat state ideal for the test of quantum gravity related energy decoherence \cite{Kha16} and would be ideal for increasing the precision of Sr atomic clocks.

\vspace{0.5cm}

{\it Acknowledgments.}I would like to thank Christoph Simon and Hon Wai Lau for fruitful discussions.


\vspace{0.5cm}

\begin{thebibliography}{100}

\bibitem{Joo03}E. Joos, et al. "Decoherence and the Appearance of a Classical World in Quantum Theory", Springer (2003); W. H. Zurek, Rev. Mod. Phys. {\bf 75}, 715 (2003); M. Arndt et al., Nature Physics {\bf 10}, 271 (2014).


\bibitem{Bas13}  A. Bassi, K. Lochan, S. Satin, T. P. Singh, and H. Ulbricht, Rev. Mod. Phys. {\bf 85}, 471
(2013).
\bibitem{Gre90}  Greenberger, et al., Am. J. Phys. {\bf 58}, 1131 (1990); Sanders, B. C. Phys. Rev. A {\bf 45}, 6811 (1992); Wenger, J., et al., Phys. Rev. A {\bf 67}, 012105 (2003); Jeong, H., et al., Phys. Rev. A {\bf 67}, 012106 (2003); Stobinska, et al., Phys. Rev. A {\bf 75}, 052105 (2007).
\bibitem{Mun02}  W. Munro et al., Phys. Rev. A {\bf 66}, 023819 (2002); Leibfried, D. et al. Science {\bf 304}, 1476 (2004).
\bibitem{Ral03}  T. C. Ralph,  et al., Phys. Rev. A {\bf 68}, 042319 (2003); M. Andrew, et al., Phys. Rev. A {\bf 72}, 052335 (2005) 
\bibitem{Enk01}  S. J. Van Enk,  et al., Phys. Rev. A {\bf 64}, 022313 (2001); H. Jeong, et al. J. Phys. Rev. A  {\bf 64}, 052308 (2001); Z. Zhao, et al. Nature {\bf 430}, 54 (2004); M. Hillery, et al., Phys. Rev. A {\bf 59}, 1829 (1999).
 \bibitem{Mon96} C. Monroe, et al., Science {\bf 272}, 1131 (1996); D. Leibfried, et al., Nature {\bf 438}, 639 (2005); T. Monz et al., Phys. Rev. Lett. {\bf 106}, 130506 (2011).
 \bibitem{Bru96} M. Brune, et al., Phys. Rev. Lett. {\bf 77}, 4887 (1996); A. Auffeves, et al., Phys. Rev. Lett. {\bf 91}, 230405 (2003); T. Meunier, et al., PRL {\bf 94}, 010401 (2005); A. Ourjoumtsev, et al., Nature {\bf 448}, 784 (2007); S. Deleglise, et al., Nature {\bf 455}, 510 (2008).
\bibitem{Fri00}  J. R. Friedman, V. Patel, W. Chen, S. K. Tolpygo, and J. E. Lukens, Nature {\bf 406}, 43 (2000).
\bibitem{Vla13}  B. Vlastakis, G. Kirchmair, Z. Leghtas, S. E. Nigg, L. Frunzio, S. M. Girvin, M. Mirrahimi, M. H. Devoret, and R. J. Schoelkopf, Science {\bf 342}, 607 (2013).

\bibitem{Zei16} J. Zeiher, et. al., Nature Physics {\bf 12}, 1095 (2016).

\bibitem{Opa15}T. Opaterny, et al., PRA {\bf 91}, 053612 (2015)
\bibitem{Mus15}W. Mussel, et al.,  PRA {\bf 92}, 023603 (2015)
\bibitem{Zoller}A.~Micheli, D.~Jaksch, J.~I.~Cirac, P.~Zoller, PRA {\bf 67}, 013607 (2003)
\bibitem{Vid11}J. Vidal, et al., PRA 70, 062304 (2004)


\bibitem{Mon11}T. Monz, et al., Phys. Rev. Lett. {\bf 106}, 130506 (2011).

\bibitem{Saf09}  M. Saffman, K. M\o lmer, PRL {\bf 102}, 240502 (2009); T. Opatrn\'y, K. M\o lmer, Phys. Rev. A {\bf 86}, 023845 (2012)
\bibitem{Opa12}T. Opatrn\'y, and K. M\o lmer, Phys. Rev. A {\bf 86}, 023845 (2012).
\bibitem{Muk11}R. Mukherjee, J. Millen, R. Nath, M.P.A Jones, and T. Pohl, J. Phys. B: At. Mol. Opt. Phys. {\bf 44} 184010 (2011).
\bibitem{Mukherjee-Thesis} R. Mukherjee, PhD thesis, Technische Universit\"{a}t Dresden (2013).
\bibitem{Kha16}  M. Khazali, H. W. Lau, A. Humeniuk, C. Simon,  Phys. Rev. A {\bf 94}, 023408 (2016). 

\bibitem{Ant14}M. Antezza, et al., PRL {\bf 113}, 023601 (2014).
\bibitem{Yan15}D. Yan, et al., Phys. Rev. A 91, 023813 (2015).
\bibitem{Car14} A. Carmele, et al., NJP {\bf 16} 063042 (2014).
\bibitem{Liu18} Y. Liu, et al., Optics Express {\bf26} 12330 (2018)
\bibitem{Meystre}F. Bariani, J. Otterbach, H. Tan, P. Meystre, Phys. Rev. A {\bf 89}, 011801(R) (2014) 

\bibitem{Muk15}Mukherjee et al. Phys. Rev. Lett. {\bf 115} 040401 (2015). 
\bibitem{Mob13}S. M\"obius, et al., Phys. Rev. A {\bf 87}, 051602 (2013).

\bibitem{Rie10}M. F. Riedel, et. al, Nature  {\bf 464},  1170 (2010).
\bibitem{Li08}Y. Li, Y. Castin, and A. Sinatra; Phys. Rev. Lett. {\bf 100}, 210401  (2008).

\bibitem{Aka08}T. Akatsuka, M. Takamoto, and H. Katori, Nature Physics {\bf 4}, 954 (2008).

\bibitem{Hen13} N. Henkel, PhD thesis, Technische Universit\"{a}t Dresden (2013).
\bibitem{Bal14}J.B. Balewski, A.T. Krupp, A. Gaj, S. Hofferberth, R. L\"ow, and T. Pfau, New J. Phys. {\bf 16} 063012 (2014).
\bibitem{Hen10}N. Henkel, R. Nath, and T. Pohl, Phys. Rev. Lett. {\bf 104}, 195302 (2010).
\bibitem{Mau11}F. Maucher, N. Henkel, M. Saffman, W. Królikowski, S. Skupin, and T. Pohl, Phys. Rev. Lett. {\bf 106}, 170401 (2011).
\bibitem{Hen12}N. Henkel, F. Cinti, P. Jain, G. Pupillo, and T. Pohl, Phys. Rev. Lett. {\bf 108}, 265301 (2012).
\bibitem{Joh10} J.E. Johnson and S.L. Rolston, Phys. Rev. A  {\bf 82}, 033412 (2010).
\bibitem{Gil14}L. I. R. Gil, R. Mukherjee, E. M. Bridge, M. P. A. Jones, and T. Pohl, Phys. Rev. Lett. {\bf 112}, 103601  (2014).


\bibitem{kuzmich}Y. O. Dudin, L. Li, F. Bariani, and A. Kuzmich, Nature Phys. {\bf 8}, 790 (2012).


\bibitem{CSS}J. M. Xiaoguang, C. Wang, , F. Nori, Physics Reports {\bf 509}  89 (2011).

\bibitem{Pez09}L. Pezz\'e and A. Smerzi, Phys. Rev. Lett. {\bf 102}, 100401 (2009).

\bibitem{Bet09} I. I. Beterov, I. I. Ryabtsev, D. B. Tretyakov, and V. M. Entin,
Phys. Rev. A {\bf 79}, 052504 (2009).

\bibitem{Cryogenic} I. Ushijima, M. Takamoto, M. Das, T. Ohkubo, and H. Katori, Nature Photonics {\bf 9}, 185 (2015);
J. M. Raimond, M. Brune, and S. Haroche, Rev. Mod. Phys. {\bf 73}, 565 (2001).

\bibitem{Sch04}V. Schweikhard, I. Coddington, P. Engels, V. P. Mogendorff, and E. A. Cornell, Phys. Rev. Lett. {\bf 92}, 040404  (2004).

\bibitem{Lattice-Lifetime}M. Takamoto, H. Katori, S. I. Marmo, V. D. Ovsiannikov, V. G. Pal'chikov, PRL 102, 063002 (2009)


\bibitem{Monte-Carlo}B. M. Garraway and P. L. Knight
Phys. Rev. A {\bf 49}, 1266 (1994); K. M\o lmer, Y. Castin, and J. Dalibard, J. Opt. Soc. Am. B {\bf 10}, 524 (1993).

\bibitem{LevelMixing}T. Keating, K. Goyal, Y. Jau, G. W. Biedermann, A. J. Landahl,  I. H. Deutsch, PRA {\bf 87}, 052314 (2013)
\bibitem{LevelMixing2}R. M. W. van Bijnen, T. Pohl, 	Phys. Rev. Lett. {\bf 114}, 243002 (2015).
\bibitem{LevelMixing3}Y.Y. Jau, A. M. Hankin, T. Keating, I. H. Deutsch, G. W. Biedermann, Nature Physics {\bf 12}, 71 (2015)




\bibitem{Saffman}Thad G. Walker and M. Saffman Phys. Rev. A {\bf 77}, 032723  (2008).




\bibitem{Pfau-Molecule} J. B. Balewski,	A. T. Krupp,	A. Gaj,	D. Peter,	H. P. B{\"u}chler, R. L{\"o}w,	S. Hofferberth	and T. Pfau, Nature {\bf 502}, 664 (2013).

\bibitem{Ott-Molecule}T. Niederpr{\"u}m, O. Thomas, T. Manthey, T. M. Weber and H. Ott, Phys. Rev. Lett. {\bf 115}, 013003 (2015).

\bibitem{KhaThe}Khazali, Mohammadsadegh. "Applications of Atomic Ensembles for Photonic Quantum Information Processing and Fundamental Tests of Quantum Physics". Diss. University of Calgary (Canada), (2016). 


\bibitem{Anomalous broadening} E.A. Goldschmidt, {\it et al.}, Phys. Rev. Lett. {\bf 116}, 113001 (2016).






\bibitem{cantilever-fabrication}M. Liao, C. Li, S. Hishita, Y. Koide, J. Micromech. Microeng. {\bf 20} 085002 (2010)
\bibitem{M-frequency}F. Brennecke,T. Donner, S. Ritter, T. Bourdel, M. Kohl, T. Esslinger, Nature {\bf 450} 268 (2007)
\bibitem{Zoller-optomechanics}A. Carmele, B. Vogell, K. Stannigel, P. Zoller, New J. Phys. {\bf 16} 063042 (2014)
\bibitem{Quality-factor}Y. Tao,	J. M. Boss,	B. A. Moores, C. L. Degen, Nature Communications {\bf 5} 3638 (2014)

\bibitem{Bayer}T. Kazimierczuk, D. Fr\"ohlich, S. Scheel, H. Stolz and M. Bayer, Nature {\bf 514}, 343 (2014).
\bibitem{Kha17}M. Khazali,  K. Heshami and C. Simon, J. Phys. B: At. Mol. Opt. Phys {\bf 50}  215301 (2017).
\end{thebibliography}
\end{document}